%% file: PRD_letter_submit.tex
\documentclass[twocolumn,aps,prd,showpacs,nofootinbib,superscriptaddress,floatfix,showkeys]{revtex4-2}

\usepackage{amsfonts}
\usepackage{amsmath}
\usepackage{amssymb} 
\usepackage{bm} % For bold math (esp of lower greek letters)
\usepackage{color}
\usepackage{enumerate}
\usepackage{graphicx}  % Needs to come after graphicx
\usepackage{dcolumn}
\usepackage{hyperref}
\usepackage{lineno}
\hypersetup{
	colorlinks=true,       % false: boxed links; true: colored links
	linkcolor=blue,          % color of internal links
	citecolor=magenta,        % color of links to bibliography
	filecolor=cyan,      % color of file links
	urlcolor=magenta           % color of external links
}

\newcommand{\D}{{\rm d}}

\labelformat{section}{#1} 
\labelformat{subsection}{Section #1} 
\labelformat{figure}{Fig.~#1} 
\labelformat{subfigure}{Fig.~\thefigure#1} 
\labelformat{table}{Tab.~#1} 
\begin{document}
\title{An upper limit on cosmological chiral gravitational wave background}
\author{Mohammad Ali Gorji}
\affiliation{Cosmology, Gravity, and Astroparticle Physics Group, Center for Theoretical Physics of the Universe, Institute for Basic Science (IBS), Daejeon, 34126, Korea}
\author{Ashu Kushwaha}
\affiliation{Department of Physics, Institute of Science Tokyo, 2-12-1 Ookayama, Meguro-ku,
	Tokyo 152-8551, Japan}
%\email{kushwaha.a.ce1b@m.isct.ac.jp}
\author{Teruaki Suyama}
\affiliation{Department of Physics, Institute of Science Tokyo, 2-12-1 Ookayama, Meguro-ku,
	Tokyo 152-8551, Japan}
\begin{abstract}
	Within the standard framework in which electroweak sphaleron processes relate lepton and baryon number, we derive an upper limit on the amplitude of a chiral gravitational wave background produced prior to the electroweak epoch. This bound is independent of the production time of chiral GWs for superhorizon modes, while it becomes sensitive to the production time for subhorizon modes. For sufficiently high reheating temperatures, the bound becomes significantly more stringent than the conventional big bang nucleosynthesis constraints at frequencies above the MHz scale, thereby providing a powerful and \emph{model-independent} probe of parity-violating physics in the early Universe.
	
\end{abstract}
\maketitle

\textbf{Introduction.}~The Universe contains overwhelmingly more baryonic matter than antimatter. This matter-antimatter asymmetry is quantified by the baryon-to-photon ratio, $\eta_B \equiv n_B/n_\gamma$, where $n_B \equiv n_b - n_{\bar b}$ is the net baryon number density, $n_b$ and $n_{\bar b}$ are the baryon and antibaryon number densities,
and $n_\gamma$ is the photon number density. Observations of the cosmic microwave background (CMB) and of the light elements produced in big bang nucleosynthesis (BBN) provide~\cite{Book-Dodelson-2020}
\begin{align}
	\eta_B =
	\begin{cases}
		6.0\times 10^{-10}
		\left( \dfrac{\Omega_b h^2}{0.022} \right)
		& \qquad\text{CMB}, \\[5pt]
		(6.040 \pm 0.118)\times 10^{-10}
		& \qquad \text{BBN},
	\end{cases}
\end{align}
where $\Omega_b h^2=0.0224 \pm 0.0002$ (95\% CL) is the baryon density parameter and $h=H_0/\left(100{\rm km}\,{\rm s}^{-1}{\rm Mpc}^{-1}\right)$ is a dimensionless number which parametrizes the Hubble constant $H_0$.
The origin of this matter-antimatter asymmetry is still an unsolved problem in cosmology~\cite{1967-Sakharov,Kuzmin:1985mm,Trodden:1998ym,1999-Riotto.Trodden,2006-Cline-arXiv,2006-Alexander.Peskin.Jabbari-PRL,Davidson:2008bu,Morrissey:2012db,Canetti:2012zc,Garbrecht:2018mrp,Xing:2020ald}. Among various proposals, Ref.~\cite{2006-Alexander.Peskin.Jabbari-PRL} provides an intriguing connection between chiral gravitational waves (GWs) and the generation of matter-antimatter asymmetry in the early Universe. The idea invokes a mechanism where the enhancement of one circular polarization of the GWs sourced during inflation can lead to the creation of a lepton number asymmetry through gravitational chiral anomaly. The asymmetry in the lepton number density is converted to baryon number density through the Standard Model electroweak sphaleron processes. Thus, this scenario of baryogenesis through leptogenesis may provide an interesting way to constrain the properties of a chiral GW background in the early Universe.

The cosmological gravitational wave background (GWB) is often assumed to be unpolarized, reflecting the absence of significant parity-violating sources in the early Universe~\cite{Maggiore-GW-V1,Book-Maggiore-Vol2,2018-Caprini.Figueroa-CQG,Domenech:2021ztg,Maleknejad:2025clz}. 
When GWs are generated by parity-symmetric processes, the left- and right-handed polarization modes are produced with equal power,
$\langle {|{h}^L (\eta,\textbf{k})|}^2 \rangle =\langle {|{h}^R (\eta,\textbf{k})|}^2 \rangle$, where ${h}^{L,R} (\eta,\textbf{k})$ are the left- and right-handed helicities of GWs in Fourier space and $\eta$ is conformal time.
This condition is equivalent to a vanishing expectation value of the gravitational Chern-Pontryagin density defined as $R\tilde{R} = \frac{1}{2} \epsilon^{\mu\nu\alpha\beta} R_{\mu\nu\rho\sigma} R^{\rho\sigma}{}_{\alpha\beta}$, where $\tilde{R}^{\mu\nu}{}_{\rho\sigma} = \frac{1}{2} \epsilon^{\mu\nu\alpha\beta} R_{\alpha\beta\rho\sigma}$ is the dual of the Riemann tensor. 
In contrast, parity-violating interactions can arise in the early Universe, and when such processes source GWs, they give rise to a chiral GWB with net circular polarization. In this case, one polarization mode is amplified while the other is suppressed through amplitude birefringence \cite{1999-Lue.Wang.Kamionkowski-PRL,2003-Jackiw.Pi-PRD,2005-Alexander.Martin-PRD,2006-Alexander.Peskin.Jabbari-PRL,2009-Alexander.Yunes-PhyRept,2010-Gluscevic.Kamionkowski-PRD,Adshead:2018doq,2021-Okano.Fujita-JCAP,Daniel:2024lev}. Consequently, $\langle {|{h}^L (\eta,\textbf{k})|}^2 \rangle  \neq \langle {|{h}^R (\eta,\textbf{k})|}^2 \rangle$ and the Chern-Pontryagin density becomes nonzero, $\langle R\tilde{R} \rangle \neq 0$.

GWs interact only very weakly with matter and therefore decouple almost immediately after their production. Consequently, constraining the GWB provides a direct probe of the mechanisms that generated them in the early Universe ~\cite{1999-Maggiore-PhyRept,2004-Bisnovatyi.Kogan-CQG,2009-Sathyaprakash.Schutz-LivRevRel,Book-Maggiore-Vol2,2018-Caprini.Figueroa-CQG,Domenech:2021ztg,Mandal:2025xuc}. 
The primary objective of the current and near-future GW detectors, such as LIGO, Virgo, KAGRA, LISA, DECIGO, BBO, and PTA, is to probe the low-frequency (below 1~kHz). In addition to this frequency band, research interest in the higher frequency region has been rapidly expanding in recent years. Experimental efforts targeting high-frequency GWs (above 10 kHz) have already begun, and several instruments are presently in operation~\cite{2006-Cruise-CQG,2008-Nishizawa.etal-PRD,2012-Cruise-CQG,2017-Chou.etal-PRD,2020-Aggarwal.etal-LivingRevRel,Domcke:2022rgu,Bringmann:2023gba,Kim:2025izt,Aggarwal:2025noe}.
Furthermore, several indirect approaches have recently been proposed to constrain such high-frequency GW signals, particularly through their imprints on radio observables~\cite{2021-Domcke.Garcia-Cely-PRL,2022-Kushwaha.etal-MNRAS,2023-Kushwaha.Sunil.Shanki-IJMPD,Ito:2023nkq,Ito:2023fcr,He:2023xoh,Domcke:2024mfu,Kushwaha:2025mia,Pappas:2025zld}. 

In this \emph{Letter}, we establish a novel \emph{model-independent} constraint on a chiral GWB by exploiting the connection between gravitational chirality and baryogenesis. 
Requiring consistency with the measured baryon asymmetry yields a \emph{model-independent} upper bound on the present-day chiral GWB amplitude, valid for both subhorizon and superhorizon modes. While the bound is insensitive to the production time for superhorizon modes, it acquires a dependence on the production time for subhorizon modes. This result provides a robust and complementary probe of parity-violating physics potentially present beyond the electroweak scale, surpassing conventional cosmological bounds such as those from BBN at high frequencies and offering a powerful new avenue to test fundamental physics beyond the Standard Model. 

We follow the $(-,+,+,+)$ metric signature, Greek indices denote the spacetime and Latin indices refer to the spatial part only. We work with natural units where $\hbar = c = k_B = 1$. The primes over the quantities refer to the derivatives with respect to conformal time $\eta$.

\textbf{Chiral Gravitational Wave Background.}~We assume that a chiral GWB was present at an initial time $\eta_i$, 
treated as a free parameter lying between the reheating and the electroweak era. 
This implies that the chiral GWB must have been generated at or before $\eta_i$ (for example, during inflation). 
We do not impose any assumptions on its production mechanism. 
In our analysis, we evaluate only the baryon number generated after $\eta_i$ by the chiral GWB. 
Any baryon asymmetry that may have been produced by the chiral GWB before $\eta_i$ or by other baryogenesis mechanisms 
is intentionally excluded in order to avoid model-dependent uncertainties. 
As a result, the upper limit on the chiral GWB derived later is conservative: including those additional contributions would only strengthen our bound. 
Furthermore, the upper limit is independent of $\eta_i$ for superhorizon modes, but exhibits a dependence on $\eta_i$ for subhorizon modes.

Chiral GWB can be quantified by the Chern-Pontryagin density $R\tilde{R}$, which can be written as a topological current up to a divergence-free term. Considering transverse-traceless metric tensor perturbations $\delta^{ij}h_{ij}=0=\partial^ih_{ij}$ in the spatially flat  Friedmann-Lemaître-Robertson-Walker (FLRW) metric $\D{s}^2 = a^2(\eta) \left[ - {\D}\eta^2 +\left(\delta_{ij}+h_{ij}\right) \D{x}^i \D{x}^j\right]$, we find
\begin{align}\label{eq-RRT-FLRW-conservation}
	R\tilde{R} = \frac{1}{a^4} \partial_\eta \left(a^3J^0\right) + \partial_i J^i \,, 
\end{align}
where $J^0 = \varepsilon_{ijk} ( {h^{\prime}}^{li} \partial^k{h^{\prime}}_l{}^j + \partial^2 h_l{}^k \partial^j h^{li} )/a^3$ in which $\varepsilon_{ijk}$ is the anti-symmetric symbol with $\varepsilon_{123}=1$. Although the explicit form of $J^i$ is not needed for the present analysis,
it is given in the Appendix~\ref{appsec-rrtilde-calc} for completeness.

To constrain the chiral GWB from the baryon asymmetry, which is created from the leptogenesis mechanism during the radiation domination (RD) era, we focus on the evolution of the GWs during the RD era. We work in Fourier space $h_{ij} (\eta,\mathbf{x}) = \sum_{s}  \int \tfrac{\D^3 {\bf k}}{(2\pi)^{3}} {h}^{s}(\eta,\textbf{k}) p^s_{ij} (\hat{\textbf{k}}) e^{ i\mathbf{k} \cdot \mathbf{x} }$, where $s=L,R$ label the left- and right-handed helicities of GWs, 
and $p_{ij}^s(\hat{\textbf{k}})$ are the usual polarization tensors (see the Appendix~\ref{appsec-rrtilde-calc} for more details). Then, the time evolution of the tensor perturbations during the RD era is generically given by
\begin{align}\label{hs-evolv}
	{h}^{s} (\eta,\textbf{k}) = {h}^s_{i} (\textbf{k})  \, \mathcal{T} (k\eta) ,
\end{align}
where $\mathcal{T} (k\eta) =  \sin[k(\eta-\eta_i)]/(k\eta) + (\eta_i/\eta) \cos[k(\eta-\eta_i)]$ is the transfer function, with initial conditions ${h}^{s} (\eta_i,{\bf k}) = {h}^s_{i} (\textbf{k})$ and ${h}^{s\,\prime} (\eta_i,{\bf k}) = 0$. Notice that, ${h}^s (\eta,{\bf k})$ remains constant as long as the mode is outside of the horizon ($k\eta \ll 1$), while it decays as $1/a$ after horizon re-entry. 
To investigate the physical consequences of the chiral GWB, we assume a \emph{model-independent} initial condition such that $\langle {|{h}^L_i (\textbf{k})|}^2 \rangle \neq \langle {|{h}^{R}_{i} (\textbf{k}) |}^2 \rangle$. Using Eq.~\eqref{hs-evolv} in Eq.~\eqref{eq-RRT-FLRW-conservation} and taking ensemble avarage (see Appendix~\ref{appsec-rrtilde-calc} for details), we find
\begin{align}\label{rrtilde-final-RD-era}
	\langle R\tilde{R} \rangle 
	&=\frac{2}{a^4}  \int \D{k} \, \,k^2 {\cal K}'(k\eta)
	\left[ \mathcal{P}^R_{h,i} (k) - \mathcal{P}^L_{h,i} (k) \right] ,
\end{align}
where $\mathcal{P}^s_{h,i}(k)$ denotes the dimensionless power spectrum of GWs 
at the initial time $\eta_i$ defined by $ \langle h^{s*}_i(\textbf{k}) h^s_i(\textbf{q}) \rangle= \frac{2\pi^2}{k^3}\mathcal{P}^s_{h,i}(k) \delta (\textbf{k}-\textbf{q})$.
In deriving Eq.~\eqref{rrtilde-final-RD-era}, we have used $\mathcal{T}(k\eta_i)=1$, $\mathcal{T}'(k\eta_i)=0$, and $\langle \partial_i J^i \rangle = 0$, and we have introduced the kernel defined as 
\begin{align}
	{\cal K}(k\eta) \equiv \mathcal{T}^2 (k\eta) - \frac{1}{k^2} {\mathcal{T}'}^2 (k\eta) \,.
\end{align}
The chirality of the GWB is described by the net circular polarization parameter which at initial time is given by
\begin{align}\label{chi-eq}
	\Delta \chi_i = \frac{\mathcal{P}_{h,i}^R - \mathcal{P}_{h,i}^L}{\mathcal{P}_{h,i}^R + \mathcal{P}_{h,i}^L} ~~.
\end{align}

Using Eq.~\eqref{chi-eq} in Eq.~\eqref{rrtilde-final-RD-era} and comparing with the ensemble average of Eq.~\eqref{eq-RRT-FLRW-conservation} gives 
\begin{align}\label{jo-chi-eq}
	\langle   a^3 J^0 \rangle =2  \int \D{k}\, k^2{\cal K}(k\eta) \, \Delta\chi_i \,\mathcal{P}_{h,i} (k) ,
\end{align}
where the total dimensionless power spectrum is given by $\mathcal{P}_{h,i}(k) = \sum_{s=L,R} \mathcal{P}^s_{h,i}(k)$.

\textbf{Constraining chiral GWB from baryon asymmetry.~}The gravitational anomaly of the total lepton current $J_{\ell}^{\mu}$ is given by~\cite{Kimura:1969iwz,ALVAREZGAUME1984269,2009-Parker.Toms-Book}
\begin{align}\label{chiral-anomaly-eq}
	\nabla_{\mu} J_{\ell}^{\mu}
	=  \frac{N_{R-L}}{384 \pi^2} R \tilde{R}~~,
\end{align}
where $N_{R-L}\equiv N_R-N_L$ is the net number of right-handed minus left-handed Weyl fermions~\cite{Maleknejad:2014wsa,Caldwell:2017chz,Adshead:2017znw,Kamada:2019ewe,Kamada:2020jaf,Maroto:2022xrv,Maleknejad:2024vvf,Mavromatos:2024szb}. The l.h.s. of the above equation describes the chirality of the fermions (or leptons), while the r.h.s. encodes the chirality of the spacetime by GWs. In that sense, the chiral GWB would act as a biased background for the evolution of fermions via the Dirac equation; as a consequence, one fermion chirality is dominant over the other. The magnitude of the generated lepton asymmetry can be determined by the comoving cosmological observer with four-velocity $u^{\mu} = (1/a,0)$ by defining the number density of chiral fermions as $n_\ell =-u_{\mu} J^{\mu}_{\ell}= a\,J^0_{\ell}$. 
\begin{figure}[t!]
	\centering
	\includegraphics[width=1.1\linewidth, height=0.3\textheight]{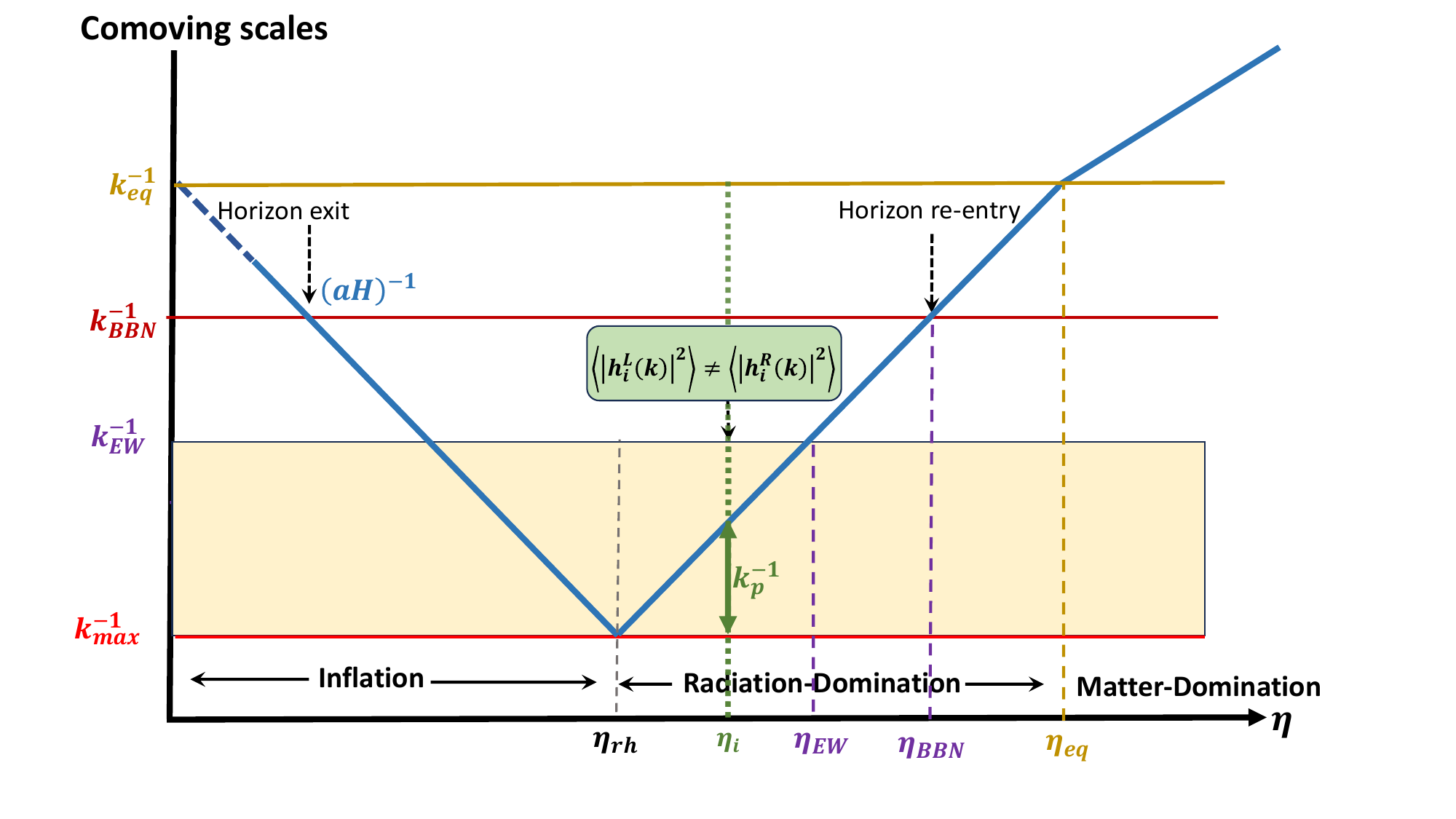}
	\caption{The figure shows the comoving scales relevant for leptogenesis. The shaded region indicates the epoch during which electroweak sphalerons efficiently convert a lepton asymmetry into a baryon asymmetry. The conformal times $\eta_{\rm rh}$ and $\eta_{\rm EW}$ mark the end of reheating and the electroweak epoch, respectively, and together define the maximally allowed region, while $\eta_{\rm BBN}$ and $\eta_{\rm eq}$ indicate BBN and matter–radiation equality, respectively. For a GW mode with comoving wavenumber $k_p$, the initial time is set by horizon crossing, \(k_p \eta_i = 1\), for super-horizon modes, while $\eta_{\rm rh}\lesssim\eta_i\lesssim\eta_{\rm EW}$ is a free parameter for the subhorizon production, e.g. $\eta_i \gg k_p^{-1}$.}
	\label{fig:scales}
\end{figure}

The link between leptogenesis and baryogenesis is provided by Standard Model electroweak sphaleron processes, which occur at temperatures higher than the electroweak scale $T_{\rm EW}\sim 100 \, {\rm GeV}$. Therefore, the creation of matter-antimatter asymmetry in the Universe through leptogenesis would be efficient only between the epoch of the end of reheating to the electroweak scales $\eta _{\rm rh} \lesssim \eta \lesssim \eta_{\rm EW} $. This is shown in \ref{fig:scales}. Once the lepton asymmetry is produced, it is converted to the baryon asymmetry as \cite{Kuzmin:1985mm,Harvey:1990qw} 
\begin{align}
	n_B = - \frac{28}{79} \, \epsilon_\ell \, n_\ell ,
\end{align}
where we have introduced an additional efficiency factor $\epsilon_\ell\leq1$, which is usually fixed to $\epsilon_\ell = 1$. Recent studies have conjectured that the total lepton asymmetry may receive contributions from vacuum effects\footnote{In the case of standard gauge interactions, one typically expects $\epsilon_\ell = \mathcal{O}(1)$ (for example, $\epsilon_\ell = 1$ for U(1) and $\epsilon_\ell = 1/6$ for SU(2))~\cite{Kamada:2024tbc}. In contrast, for the gravitational case there is currently no explicit computation of the corresponding efficiency factor. We therefore treat $\epsilon_\ell$ as a free parameter to account for this uncertainty.}, leading to $\epsilon_\ell<1$ \cite{Kamada:2024tbc}. We therefore leave $\epsilon_\ell$ unfixed to maintain flexibility in the setup.

From Eq.~\eqref{eq-RRT-FLRW-conservation} and Eq.~\eqref{chiral-anomaly-eq}, we obtain
\begin{align}
	\partial_\eta \langle  a^3 n_{\ell} \rangle  =  \frac{N_{R-L} \, }{384 \pi^2} \partial_\eta \langle   a^3 J^0 \rangle ~, %\frac{\partial}{\partial \eta}   \left( \int_0^\infty dk  ~   k^2 \left( \mathcal{T}^2 (k,\eta) - \frac{1}{k^2} {\mathcal{T}'}^2 (k,\eta) \right) \, \cdot\Delta\chi \cdot \mathcal{P}_{h,0}  \right) ~~.
\end{align}
which can be integrated from the epoch $\eta_i$ to $\eta_{\rm EW}$ as
\begin{align}\label{nl-eq}
	\left. \langle a^3 n_{\ell} \rangle \right|_{\eta_{EW}} &=  \frac{N_{R-L}}{384 \pi^2} \left( \langle a^3 J^0 \rangle|_{\eta_{EW}} -  \langle a^3 J^0 \rangle|_{\eta_i} \right) \nonumber \\
	&+ \langle a^3 n_{\ell} \rangle|_{\eta_i} ~~.
\end{align}
The last term $\langle a^3 n_{\ell} \rangle|_{\eta_i}$ characterizes any nonvanishing lepton number that exists prior to $\eta_i$. Such a lepton asymmetry may have been produced during inflation~\cite{Alexander:2004us,2006-Alexander.Peskin.Jabbari-PRL,Adshead:2017znw,Caldwell:2017chz,Kamada:2019ewe,Kamada:2020jaf} or after inflation~\cite{Fukugita:1986hr}. Moreover, there may exist theoretical mechanisms other than the sphaleron mechanism for the production of baryons, which we denote by $\eta_B^{\rm other}$. In any case, a nonzero contribution to the final total baryon number would only strengthen the resulting bound. 
As stated at the beginning of this section, we adopt the most conservative choice $\left. \langle a^3 n_{\ell} \rangle \right|_{\eta_i} = 0$ and $\eta_B^{\rm other} = 0$.

From Eqs.~\eqref{jo-chi-eq} and \eqref{nl-eq}, and taking $N_{R-L}=-3$ for the Standard Model left-handed neutrinos gives 
\begin{align}\label{etaB-fin}
	\eta_B &= \frac{7 }{1264 \pi^2} \Big( \frac{\epsilon_\ell }{n_{\gamma,0}} \Big) \int \D{k}\,  k^2 \left[ {\cal K}(k\eta_{\rm EW}) - 1\right] \,\Delta\chi_i \, \mathcal{P}_{h,i} (k) ~,
\end{align}
where $n_{\gamma,0} = a^3 n_\gamma \simeq 410.7 \, {\rm cm}^{-3}$ is the photon number density at the present epoch. Since the kernel ${\cal K}$ scales as $a^{-2}$, the dominant contribution comes from $\eta_i$, where ${\cal K}(k\eta_i)=1$. We therefore neglect ${\cal K}(k\eta_{\rm EW})$ in what follows. 

To obtain typical constraints on a chiral GWB, for simplicity, we consider a monochromatic dimensionless power spectrum sharply peaked at $k=k_p$,
\begin{align}\label{peaked-ps}
	\mathcal{P}_{h,i} (k) = \mathcal{A}_{h,i} \,  k_p \,  \delta (k-k_p) ,
\end{align}
where $\mathcal{A}_{h,i}$ is the initial amplitude at $\eta_i$. 
Substituting the above power spectrum in Eq.~\eqref{etaB-fin} and requiring that the resulting magnitude of $\eta_B$ not exceed the observed value, we obtain\footnote{
	The $k_p^{-3}$ scaling of $\eta_B$ in Eq.~\eqref{etaB-constraint-final-eq} is a generic feature of peaked primordial tensor spectra, and is not an artifact of the monochromatic ansatz. For instance, the same scaling is obtained for a lognormal spectrum $\mathcal{P}_{h,i}(k)=\frac{\mathcal{A}_{h,i}}{\sqrt{2\pi}\,\Delta}\,\exp\!\big[-\frac{\ln^2(k/k_p)}{2\Delta^2}\big]$, where $k_p$ and $\Delta$ set the location and width of the peak, respectively.
}
\begin{align}\label{etaB-constraint-final-eq}
	\mathcal{A}_{h,i} \lesssim 10 \left| \epsilon_\ell\, \, \Delta\chi_i \right|^{-1}\left( \frac{\eta_{B}^{\rm obs}}{6\times10^{-10} } \right)  \left( \frac{{10^7 \,\rm Hz}}{k_p } \right)^3~~.
\end{align}
The above equation provides an upper limit on the parameter combination $\left|\epsilon_\ell\,\Delta\chi_i\,\mathcal{A}_{h,i}\right|$ for a given $k_p$. 
\begin{figure*}[ht!]
	\centering
	\includegraphics[width=1.5\columnwidth]{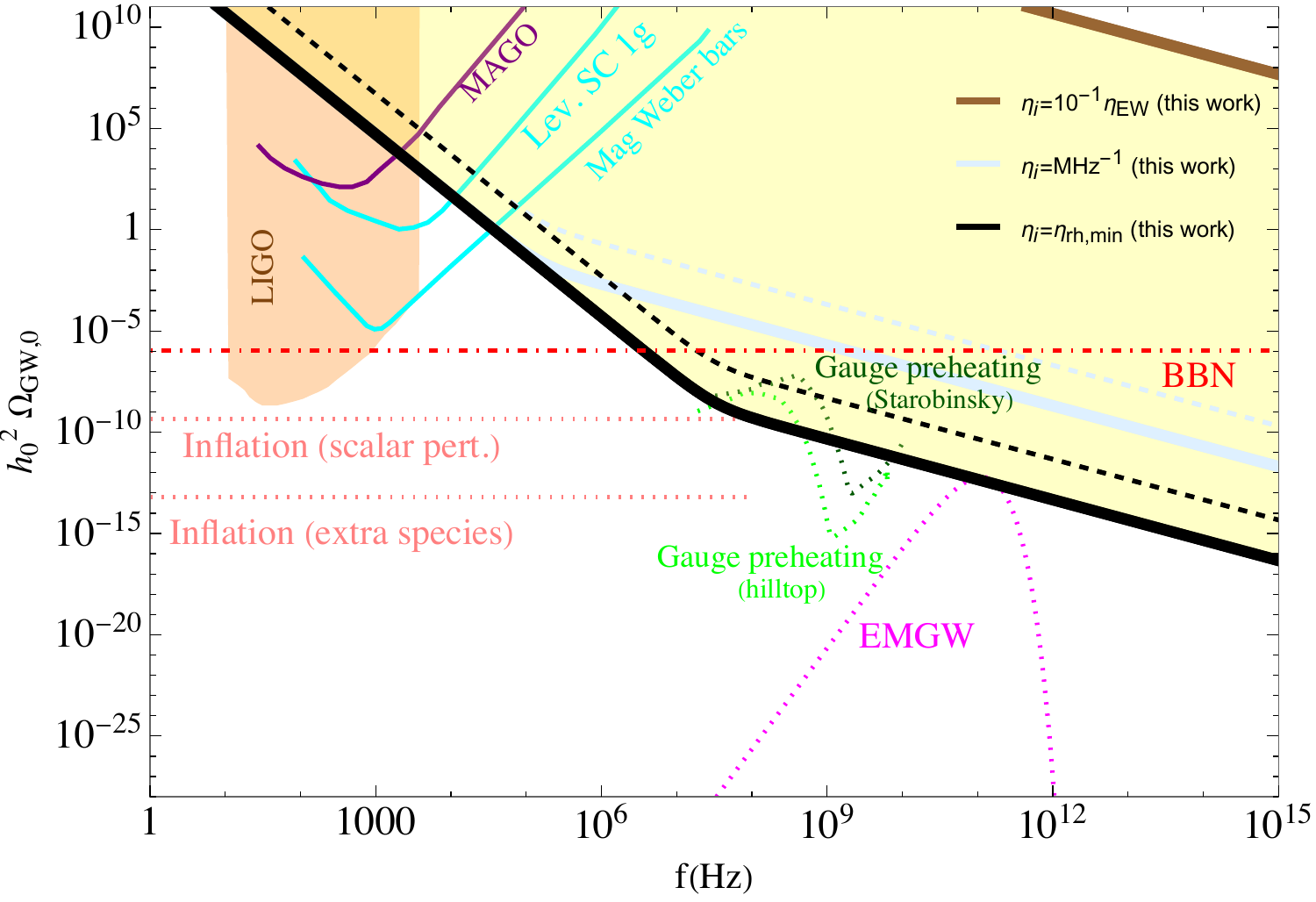}
	\caption{Constraints on total energy density of maximally chiral GWB at the present epoch by using Eq.~\eqref{gw-constraint-eq}, solid (dashed) curves for most (partially) efficient case. For comparison, we also show various detectors in high-frequency ranges (reproduced from~\cite{Aggarwal:2025noe}); Orange shaded region, purple and cyan curves are for the current experiment (published results), active R\&D effort, and proposed experiments (concept), respectively. In this figure, the horizontal axis refers to present-day frequency $f$, which corresponds to the comoving peak frequency in our case, $f_p=k_p/2\pi$. 
		As in \ref{fig:scales}, the yellow shaded region denotes the maximally allowed parameter space where our constraints are applicable. The black curve correspond to production time $\eta_i=\eta_{\rm rh,min}\simeq4.8\times10^{-9}\mathrm{Hz}^{-1}$, where $\eta_{\rm rh,min}$ refers to the time of reheating with maximally allowed reheating temperature, $T_{\rm rh,max}\simeq10^{15}\,\mathrm{GeV}$. The light-blue and  brown curves correspond to $\eta_i\simeq 10^{-6}\mathrm{Hz}^{-1}$ ($T\sim10^{12}\mathrm{GeV}$) and $\eta_i=10^{-1} \, \eta_{\rm EW}\simeq4.8\times10^{3}\mathrm{Hz}^{-1}$ ($T\sim10^3\,\mathrm{GeV}$).
		The dotted curves denote representative sources (see main text). In particular, gauge preheating and photon-graviton conversion (EMGW), shown by the green and magenta dotted curves, correspond to very high reheating temperatures and are therefore constrained by the black solid line. See Refs.~\cite{Adshead:2019igv,2024-Kushwaha.Jain-PRD} for details of the corresponding parameter space. }
	\label{fig:main-constraints}
\end{figure*}
More interestingly, we translate the above bound~\eqref{etaB-constraint-final-eq} into a constraint on the GW energy density. The energy density of GWB at $\eta_i$ is given by $\rho_{{\rm GW},i} = \int \D\ln{k} \, \frac{\D \rho_{{\rm GW},i}}{\D\ln{k}}$ which decays as $\rho_{\rm GW}\propto a^{-4}$. For the modes of our interest, we have the relation between the initial dimensionless power spectrum during the RD epoch and total energy density at the present epoch as $\Omega_{{\rm GW},0} =\rho_{{\rm GW},0}/\rho_{c,0}$, which gives,
\begin{align}
	h^2 \Omega_{{\rm GW},0} \simeq 10^{-6} \, (1+k_p^2 \eta_i^2 \,)  \,\mathcal{A}_{h,i} ,
\end{align}
where we have ignored corrections due to changes in the relativistic degrees of freedom. 
For superhorizon modes ($k_p \eta_i \ll 1$), the above relation becomes independent
of $\eta_i$, while for subhorizon modes ($k_p \eta_i \gg 1$), the r.h.s. acquires an explicit dependence on the production time $\eta_i$.
Strictly speaking, a GWB is chiral whenever the two helicities have unequal power, i.e. $\Delta\chi_i\neq 0$, and maximal chirality corresponds to $|\Delta\chi_i|=1$.
For simplicity and to obtain benchmark constraints (maximizing the observable effects), we focus on the maximally helical case $\Delta\chi_i=1$. Thus, the constraints \eqref{etaB-constraint-final-eq} implies
\begin{align}\label{gw-constraint-eq}
	h^2 \Omega_{{\rm GW},0} \lesssim 50 \epsilon_\ell^{-1} \, (1+ (2\pi f_p \eta_i)^2 \,)  \,\left( \frac{\eta_{B}^{\rm obs}}{6\times 10^{-10} } \right)  \left( \frac{{10 \,\rm kHz}}{f_p } \right)^3 ,
\end{align}
where $f_p \equiv \frac{k_p}{2\pi a_0}$ denotes the present-day peak frequency, with the scale factor normalized to unity today ($a_0 = 1$). As we can see, for the superhorizon modes, the constraint scales as $f_p^{-3}$ whereas for the subhorizon modes it scales as $f_p^{-1}$.
This is our main result: the above equation yields an upper bound on the present-day energy density (amplitude) of a chiral GWB. In \ref{fig:main-constraints}, we show our constraint: the solid (dashed) curve corresponds to the more (less) efficient case, $\epsilon_\ell=1$ ($\epsilon_\ell=0.01$), alongwith the BBN bound\footnote{Note that comparable constraints than the BBN bound have been derived from CMB observations in Refs.\cite{Smith:2006nka,Sendra:2012wh,Pagano:2015hma}, yielding $h^2 \Omega_{\rm GW,0} < 1.7\times 10^{-6}$.}, $h^2 \Omega_{\rm GW,0} < 1.1\times 10^{-6} $~\cite{Aggarwal:2025noe}. We can see that, for high reheating temperature ($T_{\rm rh}\sim 10^{15} \, {\rm GeV}$) and 
$\eta_i =\eta_{\rm rh}$, our constraint is stronger towards the high-frequency and becomes much stronger than the BBN bound above $\mathcal{O}(\rm MHz)$.

We compare our contraints with current limits from LIGO (orange shaded region), and sensitivity curves from active experiments in Research \& Development phase (purple) such as Microwave Apparatus for Graviational waves Observation (MAGO), and proposed detector concepts (cyan color) such as Magnetic Weber Bars, levitating superconducting sphere (Lev. SC) of 1g. We also discuss typical sources of chiral GWs in the early Universe (dotted lines in \ref{fig:main-constraints}). For instance, in the presence of non-Gaussian scalar perturbations, chiral GWs can be generated via a non-vanishing connected four-point function, which contributes to scalar-induced GWs \cite{Domenech:2021ztg}; alternatively, they can arise through nonlinear scalar-tensor mixing with chiral tensor perturbations \cite{Bari:2023rcw}. Chiral GWs sourced by extra fields coupled to the inflaton have also been widely studied and can yield sizable chirality \cite{Sorbo:2011rz,Maleknejad:2011jw,Maleknejad:2016qjz,Thorne:2017jft,Aoki:2025uwz,Garriga:2025uko}. 
These sources are shown in pink dotted curves. Furthermore, to illustrate the phenomenological implications of our bound, we consider two representative sources of chiral GWs in the GHz frequency range. These include chiral GWs generated through photon-graviton conversion (EMGW)~\cite{2024-Kushwaha.Jain-PRD}, shown by the magenta dotted curve, and chiral GWs produced during gauge field driven preheating after inflation~\cite{Adshead:2019igv}, depicted by the light and dark green dotted curves. The light and dark green curves correspond to the hilltop and Starobinsky inflationary models, respectively. Note that for the Starobinsky case (also including chaotic, monodromy, and D-brane models), $h^2\Omega_{\rm GW,0}$ exceeds that of the hilltop model~(see Ref~\cite{Adshead:2019igv}) and is therefore already tightly constrained by our bound. While these scenarios have previously been shown to be consistent with conventional BBN bound, if the resulting GWB is chiral our analysis demonstrates that the requirement of avoiding an excessive baryon asymmetry imposes additional and more stringent restrictions above $\mathcal{O}(\rm MHz)$ frequency range. Consequently, our results provide a complementary and tighter probe of the viable parameter space of such chiral GW production mechanisms.

\textbf{Discussion.~}A chiral GWB provides a distinctive imprint of parity-violating dynamics in the early Universe through amplitude birefringence. In this work, we studied the cosmological consequences of a chiral GWB. In the presence of a nonzero $\langle R\tilde R\rangle$, the gravitational chiral anomaly can generate a net lepton number, which is subsequently reprocessed into the observed baryon asymmetry by Standard Model electroweak sphaleron processes. Requiring consistency between the anomaly-induced contribution and the observed baryon-to-photon ratio from CMB and BBN, we derived a \emph{model-independent} upper bound on the present-day amplitude of a chiral GWB. 
The bound exhibits a distinct dependence on the production time $\eta_i$: it is independent of the $\eta_i$ for superhorizon modes, while remaining sensitive to $\eta_i$ for subhorizon modes. For maximum allowed reheating temperatures, this constraint becomes significantly stronger than the standard BBN bounds at frequencies above the MHz, opening a new observational window onto high-frequency chiral GWs. Our result therefore, provides a complementary and powerful probe of parity-violating physics beyond the Standard Model. Finally, if the observed baryon asymmetry is fully generated by mechanisms unrelated to gravitational chirality prior to the electroweak epoch, the same consistency requirement excludes any appreciable chiral GWB, making our bound maximally constraining.\\

\textbf{Acknowledgements.~}We thank Kohei Kamada and Jun'ya Kume for the discussion and clarification on their work.
The work of A.K. was supported by the Japan Society for the Promotion of Science (JSPS) as part of the JSPS Postdoctoral Program (Grant Number: 25KF0107).
The work of M.A.G. was supported by IBS under the project code IBS-R018-D3. 
The work of T.S. was support by JSPS KAKENHI grant (Grant Number JP23K03411).
M.A.G. thanks Institute of Science Tokyo for hospitality and support when this work was in progress.

\onecolumngrid
%\newpage
\appendix

\section{Computation of $\langle R\tilde R\rangle$}
\label{appsec-rrtilde-calc}

In this appendix, we provide details of the computation of $R\tilde R$ for transverse-traceless metric tensor perturbations up to quadratic order, and then evaluate its ensemble average, $\langle R\tilde R\rangle$.

\subsection{$R\tilde R$ as a total divergence}

The gravitational Chern-Pontryagin density is defined as
\begin{align}\label{RRT}
	R\tilde{R} &= \frac{1}{2} \epsilon^{\mu\nu\alpha\beta} R_{\mu\nu\rho\sigma} R^{\rho\sigma}{}_{\alpha\beta}~~;
	\qquad \tilde{R}^{\mu\nu}{}_{\rho\sigma} = \frac{1}{2} \epsilon^{\mu\nu\alpha\beta} R_{\alpha\beta\rho\sigma} ,
\end{align}
where $\epsilon^{\delta\eta\alpha\beta}$ is the Levi-Civita tensor, with $\epsilon^{0123}=-1/\sqrt{-g}$. The Chern-Pontryagin density can be written as a total divergence,
\begin{align}
	R{\tilde R} = 2 \nabla_\mu K^\mu \,,
\end{align}
where the current $K^\mu$ is not unique and is defined up to the addition of a divergence-free term. One of the most well-known choices for the current is
\begin{align}\label{K-def}
	K^\mu = \epsilon^{\mu\eta\alpha\beta} \left(
	\Gamma^\delta_{\eta\lambda} \partial_\beta \Gamma^\lambda_{\alpha\delta} 
	+ \frac{2}{3} \Gamma^\delta_{\eta\sigma} \Gamma^\sigma_{\beta\lambda} \Gamma^\lambda_{\alpha\delta} 
	\right) \,,
\end{align}
where $\Gamma^{\rho}_{\mu\nu}$ is the Christoffel connection.

Considering transverse-traceless metric tensor perturbations $\delta^{ij}h_{ij}=0=\partial^ih_{ij}$ in the spatially flat FLRW background
\begin{align}\label{metric}
	\D{s}^2 = - {\D}t^2 + a^2(t) \left(
	\delta_{ij}+h_{ij}\right) \D{x}^i \D{x}^j \,,
\end{align}
where $t$ denotes the cosmic time, 
up to quadratic order in perturbations, we find
\begin{align}\label{RRT-FLRW}
	R{\tilde R} = 
	\frac{2}{a}\,\varepsilon_{bcd}
	\left[
	\big(\ddot h^{ab}+H\,\dot h^{ab}\big)\,\partial^d \dot h_{a}{}^c
	\;+\;
	\frac{1}{a^2} \big(\partial_a \dot h_{e}{}^d-\partial_e \dot h_{a}{}^{d}\big)\,\partial^e\partial^c h^{ab}
	\right] \,,
\end{align}
where a dot denotes $\D/\D{t}$ and $H=\dot{a}/a$ is the Hubble parameter. Here $\varepsilon_{ijk}$ is the anti-symmetric symbol with $\varepsilon_{123}=1$.

Substituting the metric Eq.~\eqref{metric} in Eq.~\eqref{K-def} and expanding up to the second order in perturbations, we find
\begin{align}\label{eq:K0}
	K^0 &= 
	\frac{1}{2a}\,\varepsilon_{b c d}
	\left[
	\dot h^{a b}\,\partial^{d}\dot h_{a}{}^{c}
	\;+\; \frac{1}{a^2} \partial_{e} h^{a b} \partial^{d}\left(
	\partial_{a} h^{e}{}^{c}
	- \partial^{e} h_{a}{}^{c} \right)
	\right] \,,
	\\ \label{eq:Ki}
	K^i &= -\;\frac{1}{2a}\,\varepsilon^i{}_{b c}\,
	\left[
	\dot h^{a b} \ddot h_{a}{}^{c}
	\;+\; \frac{1}{a^2} 
	\partial_{d} h^{a b}
	\big(
	\partial_{a}\dot h^{d}{}^{c}
	-
	\partial^{d}\dot h_{a}{}^{c}
	\big)
	\right] \,.
\end{align}

Since the Chern-Pontryagin density \eqref{RRT} is a total derivative, in the FLRW case it can be written as
\begin{align}\label{RRT-FLRW-conservation}
	\tilde{R}R = \frac{1}{a^3} \partial_t \left(a^3J^0\right) + \partial_i J^i \,.
\end{align}
A natural choice for $J^0$ and $J^i$ is $J^0=2K^0$ and $J^i=2K^i$, which is commonly adopted in the cosmology literature. However, this choice is not unique. For example, one can show that
\begin{align}\label{eq:J0}
	J^0 &= \frac{1}{a}\,\varepsilon_{b c d}
	\left(
	\dot h^{a b}\,\partial^{d}\dot h_{a}{}^{c}
	\;+\;
	\frac{1}{a^2}\partial^{c} h^{a b}\,
	\partial^2 h_{a}{}^{d}
	\right) \,,
	\\ \label{eq:Ji}
	J^i &= -\frac{1}{a}\,\varepsilon_{bc}{}^i\,\dot h^{a b}
	\Big(
	\ddot h_{a}{}^c
	-\frac{1}{a^2}\,\partial^2 h_{a}{}^{c}
	\Big)
	+\frac{1}{a^{3}}\varepsilon_{b c d}
	\left[ \partial^{i}\!\big(\dot h^{a b}\,\partial^{c} h_{a}{}^d\big)
	-2\partial^{a}\dot h^{i b}\,\partial^{c} h_{a}{}^{d}
	\right] \,,
\end{align}	
also satisfies \eqref{RRT-FLRW-conservation} and provides an alternative choice of current.

Comparing the above result with \eqref{eq:K0} and \eqref{eq:Ki}, we see that the temporal component $J^0$ is simpler in \eqref{eq:J0}, while the spatial component $J^i$ is more involved. Since in this paper we only make use of $J^0$, we work with \eqref{eq:J0} and \eqref{eq:Ji}, which are more convenient for practical computations in cosmology.

\subsection{Computational details for $\langle R\tilde R\rangle$}
\label{appsec-rrtilda-details}
Working with conformal time $\D\eta=\D{t}/a(t)$, Eq.~\eqref{RRT-FLRW} takes a simpler form
\begin{align}\label{rr-tilde-eta-main}
	R\tilde{R} = \frac{2}{a^4}  \varepsilon_{ijk} \left[  {h^{\prime\prime}}^{li} \partial^k {h^{\prime}}^j_l + \left( \partial_l {h^{\prime}}^k_m - \partial_m {h^{\prime}}^k_l  \right) \partial^m \partial^j h^{li} \right] \,,
\end{align}
where a prime denotes $\D/\D\eta$ and Eq.~\eqref{RRT-FLRW-conservation} takes the form
\begin{align}\label{ap-eq-RRT-FLRW-conservation}
	R\tilde{R} = \frac{1}{a^4} \frac{\partial}{\partial \eta} \left(a^3J^0\right) + \partial_i J^i \, \,. 
\end{align}
Let us consider the Fourier space decomposition of $h_{ij} (\eta,\textbf{x})$ in terms of the helicities as
\begin{align}\label{eq:hij-lr}
	h_{ij} (\eta,\mathbf{x}) = \sum_{s=L,R}  \int \frac{\D^3 \textbf{k}}{(2\pi)^{3}} {h}^{s} (\eta,\textbf{k}) \, p^s_{ij} (\hat{\textbf{k}}) e^{ i\mathbf{k} \cdot \mathbf{x} } \,.
\end{align}
where we note that the Fourier amplitudes ${h}^{s} (\eta,\textbf{k})$ are complex quantities. The two polarization tensors are real, $p_{ij}^s (-\hat{\textbf{k}}) = p_{ij}^s (\hat{\textbf{k}}) $ and the condition for $h_{ij} (\eta, \textbf{k})$ to be real can be given as ${h}^*_{ij} (\eta, \textbf{k}) = {h}_{ij} (\eta, -\textbf{k})$. The two polarization tensors depend on the unit vector $\hat{\textbf{k}}$ and satisfy the following properties $p^s_{ij} = p^s_{ji}$ (symmetric), $\hat{k}_i p^s_{ij} = 0$ (transverse), and $p^s_{ii} = 0$ (traceless). The polarization tensors satisfy the following relations
\begin{align}\label{ortho-rel}
	p^R_{ij} (\hat{\textbf{k}}) p^R_{ij} (\hat{\textbf{k}}) &= 0=  p^L_{ij} (\hat{\textbf{k}}) p^L_{ij} (\hat{\textbf{k}}) , \\
	p^R_{ij} (\hat{\textbf{k}}) p^L_{ij} (\hat{\textbf{k}}) &= 2 , \\
	\label{comp-rel}
	k_p \varepsilon^{mpj} p^s_{ij} (\hat{\textbf{k}}) &= - i \lambda_k^s \, k \,  {p^m_i}^s (\hat{\textbf{k}}),
\end{align}
where $s=R,L$, $\lambda_k^R = +1$, and  $\lambda_k^L=-1$. The helicity relations of the polarization tensors for the circularly polarized GWs (left-handed and right-handed) can be defined in terms of the linearly polarized polarization tensors as
\begin{align}
	p^R_{ij} = \frac{p^+_{ij} + i\, p^{\times}_{ij}}{\sqrt{2}} \, , \quad p^L_{ij} = \frac{p^+_{ij} - i\, p^{\times}_{ij}}{\sqrt{2}} ~~.
\end{align}
The volume average or ensemble average of Eq.~\eqref{ap-eq-RRT-FLRW-conservation} is given by
\begin{align}\label{rrtilde-j-rel}
	\langle R\tilde{R} \rangle = \Big\langle \frac{1}{a^4} \partial_{\eta} (a^3 J^0) \Big\rangle + \langle \partial_i J^i \rangle = \frac{1}{a^4} \Big\langle  \partial_{\eta} (a^3 J^0) \Big\rangle \,,
\end{align}
where in the last equality we used the fact that in homogeneous and isotropic FLRW universe, $\langle \partial_i J^i \rangle=0$. Furthermore, since $\langle \cdots\rangle$ is an ensemble average which is related to a volume average, we can take $\partial_{\eta}$ outside the averaging
\begin{align}
	\langle R\tilde{R} \rangle = \frac{1}{a^4} \frac{\partial}{\partial \eta} \Big\langle  \varepsilon_{ijk} \left[ {h^{\prime}}^{li} \partial^k{h^{\prime}}^j_l + \partial^2 h^k_l \partial^j h^{li} \right] \Big\rangle = \frac{1}{a^4} \frac{\partial}{\partial \eta}   \left[ \Big\langle  \varepsilon_{ijk} {h^{\prime}}^{li} \partial^k{h^{\prime}}^j_l \Big\rangle +  \Big\langle  \varepsilon_{ijk} \partial^2 h^k_l \partial^j h^{li} \Big\rangle \right] .
\end{align}
In Fourier space, we can compute both terms on r.h.s. in the above equation as
\begin{align}
	\Big\langle  \varepsilon_{ijm} {h^{\prime}}^{li} \partial^m{h^{\prime}}^j_l \Big\rangle & = \Big\langle \int \frac{\D^3 \textbf{k}}{(2\pi)^{3}}  ~ \int \frac{\D^3 \textbf{q}}{(2\pi)^{3}}  ~ \sum_{r=L,R}\sum_{s=L,R} \, \varepsilon_{ijm} (i q^m) \, {h}^{r\,\prime} (\eta,{\bf k}) \, {h}^{s\,\prime} (\eta,{\bf q}) \,  
	{p^{li}}^r (\hat{\textbf{k}}) {p^j_l}^s (\hat{\textbf{q}}) e^{ i\mathbf{k} \cdot \mathbf{x} +  i\mathbf{q} \cdot \mathbf{x} } 
	\Big\rangle
	\nonumber\\ 
	&= \int \frac{\D^3 \textbf{k}}{(2\pi)^{3}}  ~ \int \frac{\D^3 \textbf{q}}{(2\pi)^{3}}  ~ \sum_{r=L,R}\sum_{s=L,R} \, \varepsilon_{ijm} (i q^m) \, {h}^{r\,\prime} (\eta,{\bf k}) \, {h}^{s\,\prime} (\eta,{\bf q}) \,  {p^{li}}^r (\hat{\textbf{k}}) {p^j_l}^s (\hat{\textbf{q}}) (2\pi)^{3} \delta^3 (\mathbf{k}+\mathbf{q}) 
	\nonumber\\ 
	&= \int \frac{\D^3 \textbf{k}}{(2\pi)^{3}}  ~  \sum_{r=L,R}\sum_{s=L,R} \, \varepsilon_{ijm} (-i k^m) \, {h}^{r\,\prime} (\eta,{\bf k}) \, {h}^{s\, \prime} (\eta,{\bf -k}) \, {p^{li}}^r (\hat{\textbf{k}}) {p^j_l}^s (\hat{-\textbf{k}}) 
	\nonumber\\ 
	&= \int \frac{\D^3 \textbf{k}}{(2\pi)^{3}}  ~  \sum_{r=L,R}\sum_{s=L,R} \, (-i^2 \lambda^s k) \, {h}^{r\,\prime} (\eta,{\bf k}) \, {h}^{s\, \prime} (\eta,{\bf -k}) \,  {p^{li}}^r (\hat{\textbf{k}}) p_{il}^s (\hat{-\textbf{k}}) 
	\nonumber\\ 
	&= 2\int \frac{\D^3 \textbf{k}}{(2\pi)^{3}}  \,  k \,  \left[ {h}^{L\,\prime} (\eta,{\bf k}) \, {h}^{R\, \prime} (\eta,{\bf -k})- {h}^{R\,\prime} (\eta,{\bf k}) \, {h}^{L\, \prime} (\eta,{\bf -k}) \right]
	\nonumber\\ 
	&= 2\int \frac{\D^3 \textbf{k}}{(2\pi)^{3}}   k \,  \left[ {h}^{L\,\prime} (\eta,{\bf k}) \, {{h}^{L\, \prime} (\eta,{\bf k})}^*- {h}^{R\,\prime} (\eta,{\bf k}) \, {{h}^{R\, \prime} (\eta,{\bf k})}^* \right]
	\nonumber\\ 
	&= 2 \int \frac{\D^3 \textbf{k}}{(2\pi)^{3}}  ~ k \,  \left[ \left| {h}^{L\,\prime} (\eta,{\bf k}) \right|^2 - \left|{h}^{R\,\prime} (\eta,{\bf k}) \right|^2 \right] ,
\end{align}
where in the second line we used the relation for the volume average $\langle e^{ i\mathbf{k} \cdot \mathbf{x} +  i\mathbf{q} \cdot \mathbf{x} }  \rangle = \int \D^3 \textbf{x} \, e^{ i\mathbf{k} \cdot \mathbf{x} +  i\mathbf{q} \cdot \mathbf{x} } = (2\pi)^{3} \delta^3 (\textbf{k}+\textbf{q})$, and we have also used ${{h}^{R} (\eta,{\bf k})}^* = {h}^{L} (\eta,{\bf -k})$ and ${{h}^{L}(\eta,{\bf k})}^* = {h}^{R} (\eta,{\bf -k})$.
Similarly, we obtain the second term as
\begin{align}
	\Big\langle  \varepsilon_{ijm} \partial^2 h^m_l \partial^j h^{li} \Big\rangle & = \Big\langle \int \frac{\D^3 \textbf{k}}{(2\pi)^{3}}  ~ \int \frac{\D^3 \textbf{q}}{(2\pi)^{3}}  ~ \sum_{r=L,R}\sum_{s=L,R} \, \varepsilon_{ijm} (-i k^2 q^j) \, {h}^{r} (\eta,{\bf k}) \, {h}^{s} (\eta,{\bf q}) \,  
	{p^m_l}^r (\hat{\textbf{k}}) {p^{li}}^s (\hat{\textbf{q}}) e^{ i\mathbf{k} \cdot \mathbf{x} +  i\mathbf{q} \cdot \mathbf{x} } 
	\Big\rangle
	\nonumber\\ 
	&= 2\int \frac{\D^3 \textbf{k}}{(2\pi)^{3}}  ~   k^3 \,  \left[ \left| {h}^{R} (\eta,{\bf k}) \right|^2 - \left|{h}^{L} (\eta,{\bf k}) \right|^2 \right]~~.
\end{align}
Therefore, using these two results give 
\begin{align}\label{appeq-rrtilde-final}
	\langle R\tilde{R} \rangle 
	&= \frac{2}{a^4} \frac{\partial}{\partial \eta} \left( \int \frac{\D^3 \textbf{k}}{(2\pi)^{3}} ~  k \,  \left[   k^2 \left( \left|{h}^{R} (\eta,{\bf k}) \right|^2 - \left| {h}^{L} (\eta,{\bf k}) \right|^2 \right) - \left( \left| {h}^{R\,\prime} (\eta,{\bf k}) \right|^2 - \left|{h}^{L\,\prime} (\eta,{\bf k}) \right|^2 \right) \right] \right) ~.
\end{align}
The evolution of GWs in the RD era is given by~\cite{Book-Maggiore-Vol2}
\begin{align}\label{ap-hs-evolv-A-B}
	h^{s} (\eta,\textbf{k}) = A (k) \frac{\sin{(k\eta)} }{k\eta} + B (k) \frac{\cos{(k\eta)} }{k\eta} \,,
\end{align}
where $A(k)$ and $B(k)$ are arbitrary constants and can be fixed by the initial conditions. Note that in the superhorizon limit $k\eta \ll 1$, the solution $\sin{(k\eta)} /(k\eta)$ refers to the constant mode and $\cos{(k\eta)} /(k\eta)$ (goes as $1/\eta$) corresponds to the decaying mode. Hence, often the term $B (k) \cos{(k\eta)}/ (k\eta)$ in Eq.~\eqref{ap-hs-evolv-A-B} is ignored while considering the evolution of GWs. But, this mode is important in our case and we work with the full solution. We choose the initial conditions as $h^{s} (\eta_i,\textbf{k}) = h^s_i (\textbf{k})$ and ${h^{s\,\prime}}(\eta_i,\textbf{k}) = 0$, which gives
\begin{align}\label{ap-hs-evolv}
	h^{s} (\eta,\textbf{k}) = h^s_i (\textbf{k}) \mathcal{T} (k\eta) 
	\quad \text{with} \quad \mathcal{T} (k\eta) =  \frac{\sin{[k(\eta-\eta_i)]} + k \eta_i \cos{[k(\eta-\eta_i)]}}{k\eta} \,,
\end{align}
where $\mathcal{T} (k\eta)$ is the transfer function. Using Eq.~\eqref{ap-hs-evolv} in Eq.~\eqref{appeq-rrtilde-final}, we obtain
\begin{align}\label{ap-rrtilde-final-RD-era}
	\langle R\tilde{R} \rangle 
	=\frac{2}{a^4} \frac{\partial}{\partial \eta}   \left( \int \frac{\D{k}}{2\pi^2} ~  k^5  \mathcal{K} (k\eta) \,  \left[ |{h}_i^R ({\bf k}) |^2 - |{h}_i^L ({\bf k}) |^2 \right]  \right) \,,
\end{align}
where we used ${h^{s\,\prime}}(\eta,\textbf{k})=h^s_i (\textbf{k}) \, \mathcal{T}' (k\eta) $ and the kernel is defined as
\begin{align}\label{ap-transfer-func-terms}
	\mathcal{K} (k\eta) \equiv \mathcal{T}^2 (k\eta) - \frac{1}{k^2} {\mathcal{T}'}^2 (k\eta) &=  \left(\frac{1}{k\eta} \right)^4 \, \Bigg[~ (k \eta)^2 \Big\{ k\eta_i \cos{[k(\eta-\eta_i)]} + \sin{[k(\eta-\eta_i)]} \Big\}^2 
	\nonumber \\ 
	&  \qquad\qquad\qquad - \Big\{ k(\eta-\eta_i) \cos{[k(\eta-\eta_i)]} - (1+k^2 \eta \eta_i ) \sin{[k(\eta-\eta_i)]} \Big\}^2 ~\Bigg] ~~.
\end{align}

\bibliographystyle{apsrev4-1}
\input{PRD_letter_submit.bbl}
%\bibliography{References}
\end{document}

%% file: PRD_letter_submit.bbl
%merlin.mbs apsrev4-1.bst 2010-07-25 4.21a (PWD, AO, DPC) hacked
%Control: key (0)
%Control: author (72) initials jnrlst
%Control: editor formatted (1) identically to author
%Control: production of article title (-1) disabled
%Control: page (0) single
%Control: year (1) truncated
%Control: production of eprint (0) enabled
%